# EFFICIENCY LIMIT OF $Al_xGa_{1-x}As$ SOLAR CELL MODIFIED BY $Al_yGa_{1-y}Sb$ QUANTUM DOT INTERMEDIATE BAND EMBEDDED OUTSIDE OF THE DEPLETION REGION

A. Kechiantz[1, 2], A. Afanasev[1], J.-L. Lazzari[3], A. Bhouri[4], Y. Cuminal[5], and P. Christol[5]
[1]Department of Physics, The George Washington University,
725 21st Street, NW, Washington, DC, 20052, USA; kechiantz@gwu.edu; afanas@gwu.edu
[2]On leave from Institute of Radiophysics and Electronics, National Academy of Sciences,
1 Brothers Alikhanyan st., Ashtarak 0203, ARMENIA; mrs_armenia@yahoo.com
[3]Centre Interdisciplinaire de Nanoscience de Marseille, CINaM, UMR CNRS 7325, Aix-Marseille Université,
Campus de Luminy, Case 913, Avenue de Luminy, 13288 Marseille Cedex 9, France ; lazzari@cinam.univ-mrs.fr
[4]Departement de Physique, Faculté des Sciences de Monastir, 5019 Monastir, Tunisia; bhouri_amel@yahoo.fr
[5]Institut d'Electronique du Sud (IES), UMR CNRS 5214, Case 067, Université Montpellier 2,
34095 Montpellier Cedex 05, France; philippe.christol@ies.uni; yvan.cuminal@ies.univ-montp2.fr

ABSTRACT: Recombination through quantum dots (QDs) is a major factor that limits efficiency of QD intermediate-band (IB) solar cells. Our proposal for a new IB solar cell based on type-II GaSb QDs located "outside" the depletion region of a GaAs p-n-junction aims to solve this problem. The important advantage of proposed heterostructure appears due to the "outside" location of IB. Such IB does not assist generation of additional leakage current flow through the depletion region. Carriers cannot escape from "outside" QDs through the buffer layer and the depletion region into GaAs substrate by tunneling because QDs are far from the depletion layer. Only solar photon or thermal assistance may enable electron escape from QDs. Such type-II QD IB solar cell concept promises an efficiency enhancement relative to that of GaAs solar cells.
Keywords: Intermediate Band Solar Cells, GaSb/GaAs type-II Quantum Dots, High-Efficiency Solar Cells.

## 1 INTRODUCTION

The intermediate band (IB) is a concept promising about to double the limiting efficiency of single-junction solar cells by putting into use energy of sub-band gap photons [1]. Such low-energy photons may generate additional photocurrent in solar cells due to the nonlinear effect of two-photon absorption that supplies a valence band electron enough energy for transition into the conduction band [2]. Concentration of the sub-band gap photon flux and introduction of intermediate electronic state in semiconductor band gap are necessary conditions for escalating the nonlinear effect by resonant two-photon absorption [2].

The IB concept does not define the structure of IB solar cell, in particular, whether IB states should be located "inside" [3] or "outside" [4] of the depletion region of p-n-junction in solar cell. Recent theoretical [4] and experimental [3, 5] studies have shown that IB states work as recombination centers. Involvement of such centers in photocurrent generation inevitably leads to additional dark current generation even in ideal IB solar cells. For instance, IB composed by type-I quantum dots (QDs) located "inside" of the depletion region degrades the open circuit voltage and the conversion efficiency of IB solar cell relative to that of the reference cell [3]. The latter cell has a similar structure but without QDs.

Concentration of sunlight suppresses the relative effect of recombination through IB states located "outside" of the depletion region [4]. Calculation has shown that about 500-sun concentration may raise the limiting efficiency of Si solar cell with Ge type-II QDs running as IB located "outside" of the depletion region, by 25% relative to that of the reference cell [2, 4]. Noteworthy both Si and Ge are indirect band gap semiconductors.

In this work we focus on GaAs/GaSb strained system of direct band gap semiconductors having a type-II misalignment of the conduction and valence bands. The large offset, direct band gaps, and well-developed fabrication technology make this material system a valuable candidate for studying and understanding of IB potential of realistic type-II QDs.

For such study we use IB solar cell heterostructure based on an absorption region built-on type II $Al_yGa_{1-y}Sb$ strained QDs stacked in an $Al_yGa_{1-y}As$ gradual direct band gap layer separated from the depletion region of a GaAs p-n-junction by a buffer layer. We have already shown that the two-photon absorption may increase the limiting efficiency of such IB solar cells by 20% relative to that of the reference GaAs solar cell [6]. Taking into account both radiative and non-radiative recombination associated with QDs, we here focus on the effect of the sunlight concentration on the photovoltaic characteristics and the efficiency.

## 2 TYPE-II QD IB SOLAR CELL

### 2.1 Layout and electronic features of the cell

Figure 1 displays the structure of new IB GaAs solar cell with GaSb QDs. The cell includes a graded epitaxial stack of $Al_yGa_{1-y}Sb$ strained QD layers alternating with undoped $Al_xGa_{1-x}As$ spacers ($0<x, y<0.40$). The stack is sandwiched between a *p*-doped GaAs cap layer and a thin *p*-doped GaAs buffer layer grown on an $n^+$-doped GaAs substrate so that the buffer and substrate compose the *p*-*n*-junction that the edge of depletion region cannot reach the first QD and spacer layers.

An important feature of this structure is that both buffer and stack are thin as compared to electron and hole diffusion lengths. On the other hand, the buffer is thick enough, so that the photoelectrons confined in type-II $Al_yGa_{1-y}Sb$ QDs cannot escape from QDs into the GaAs substrate by tunneling through the buffer.

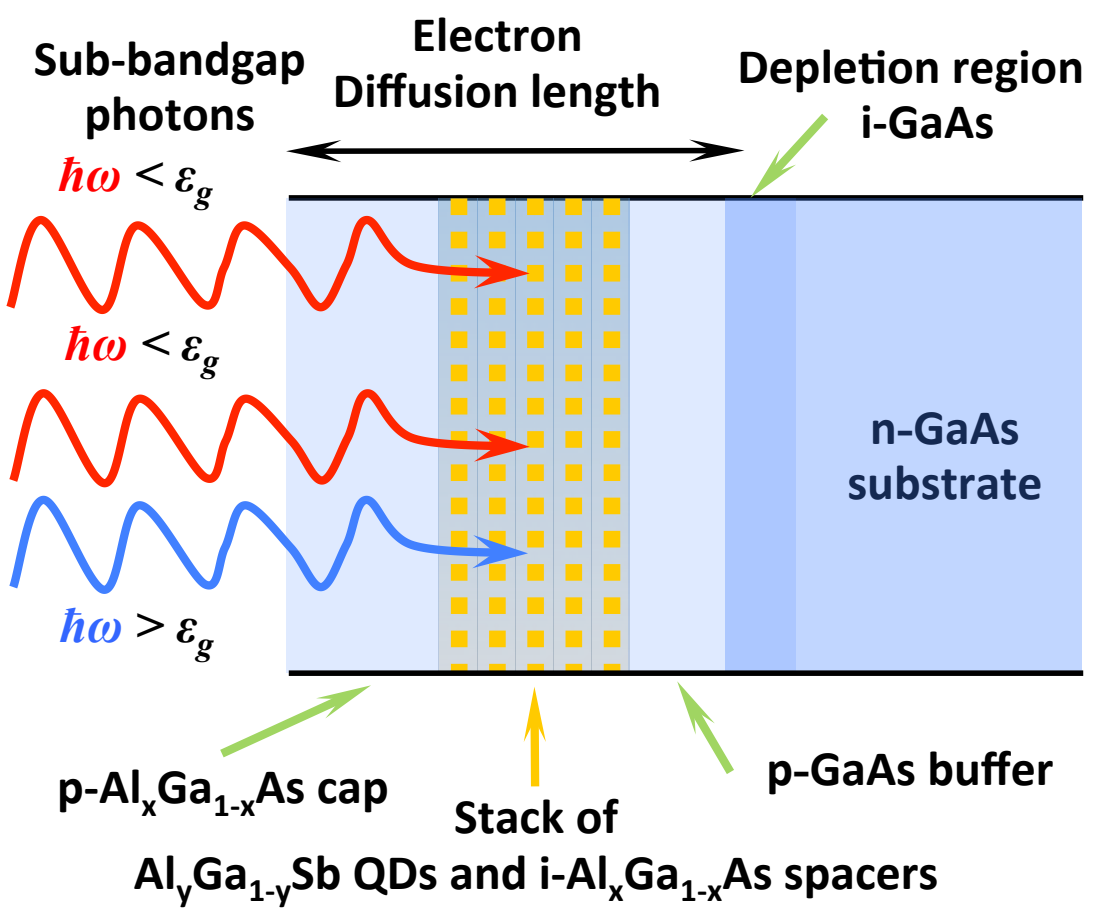

**Figure 1:** Layout of GaSb/GaAs type-II QD IB solar cell. $Al_yGa_{1-y}Sb$ QDs are embedded in the graded $Al_xGa_{1-x}As$ layer located outside of the depletion region of the GaAs p-n-junction

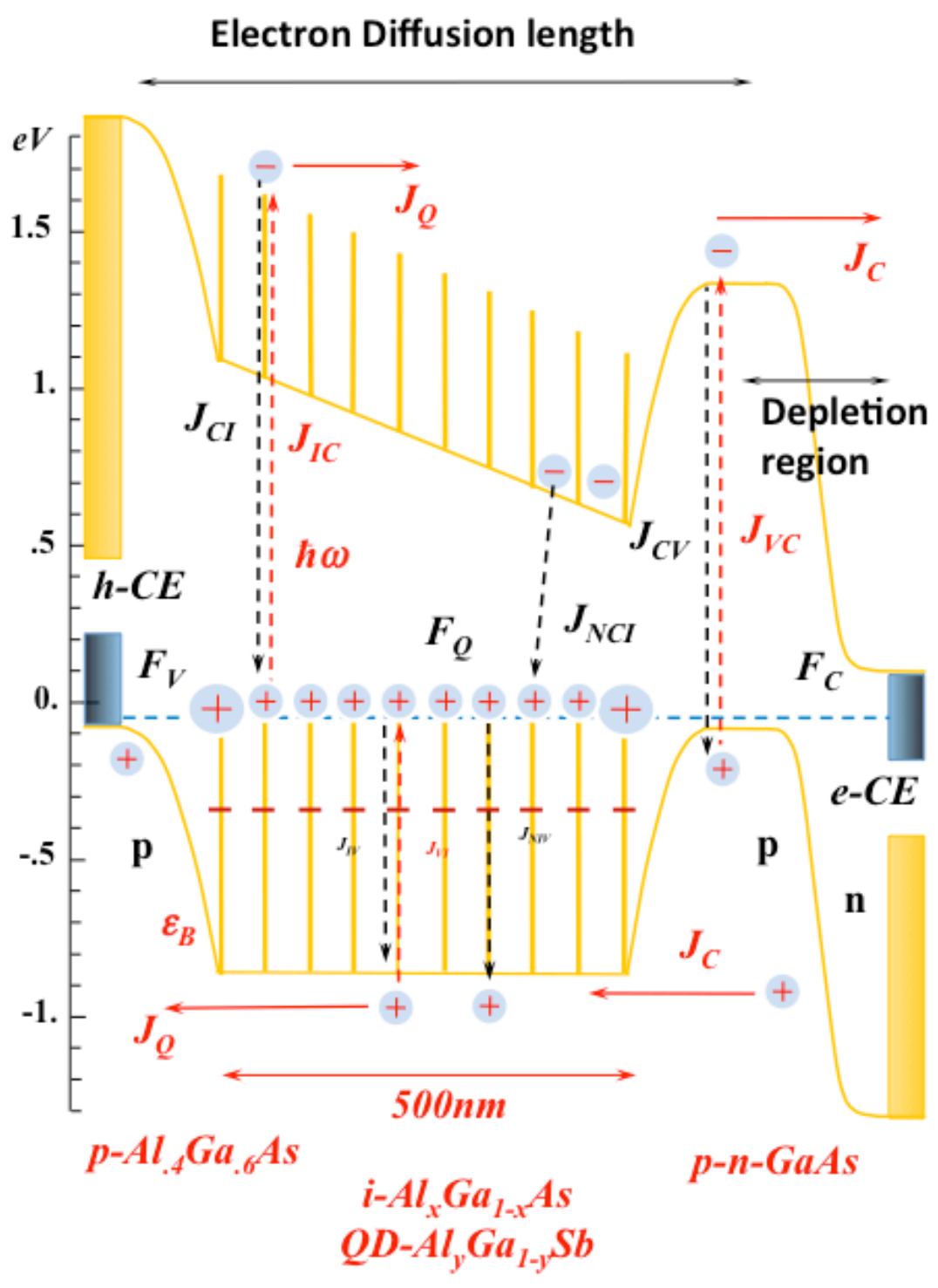

**Figure 2:** Energy band diagram of GaSb/GaAs type-II QD IB solar cell. $Al_yGa_{1-y}Sb$ QDs are embedded in graded $Al_xGa_{1-x}As$ layer located outside of the depletion region [5]; the yellow rods with the red crosses on the top are the type-II QDs; the white strips in conduction and valence bands are the $Al_xGa_{1-x}As$ spacers; the stack of spacers is 500 nm thick; $j_{VC}$, $j_{VI}$ and $j_{IC}$ are generation currents; $j_{CV}$, $j_{IV}$ and $j_{CI}$ are radiative recombination currents; $j_{NIV}$ and $j_{NCI}$ are non-radiative recombination currents; $e-CE$ and $h-CE$ are electron and hole contact electrodes

Another essential feature is modulation doping of QDs by the buffer and cap layers. For equalizing Fermi level in the cell, *p*-doped cap and buffer layers supply holes into confined electronic states in QDs. These holes bring a positive charge into QDs. The charge shifts conduction and valence band edges of the stack down relative to that of the cap and buffer layers. Figure 2 shows rearrangement of energy bands due to such charging of QDs. A barrier of $\varepsilon_B$ height separates the stack from the cap and buffer layers. Holes are still majority carriers in the stack of $Al_xGa_{1-x}As$ spacers while mobile photoelectrons are in $Al_xGa_{1-x}As$ internal field pulling them towards the buffer layer.

For instance, if the stack comprises ten epitaxial $Al_xGa_{1-x}As$ spacers each of 50 *nm* thick, varying from x=0 at the buffer to x=0.4 at the cap layer, their band gaps vary from 1.43 *eV* to 1.83 *eV*, respectively. Such variation creates about 1 *kV/cm* field pushing electrons towards the depletion region. The field is so strong that photoelectrons acquire velocity of about $10^6$ *cm/s* and pass 500 *nm* of such stack in 50 *ps*.

The type-II lineup of energy bands spatially separates photoelectrons of conduction band from holes confined in QDs. Such separation blocks mobile photoelectron wave function from that of holes confined in QDs. The reduced overlapping of wave functions slows down to 10 *ns* the $\tau_C$ lifetime associated with non-radiative electron-hole inter-band recombination through QDs [7].

The next feature of the proposed structure shown in Figure 2 is that the Fermi level lowers amid the confined electronic states of QDs. Since both electrons and holes are gathered at the Fermi level, the density of both occupied and unoccupied confined states in such QDs is as high as in heavily doped semiconductors. Since intra-band and inter-band absorption coefficients of QDs are proportional to the density of occupied and unoccupied confined electronic states in QDs, both coefficients must be about equal to that of bulk direct band gap semiconductors.

Another feature is a small, about $10^{-18}$ $cm^3$, volume of QDs. Injection of a single photoelectron into the confined electronic state raises the density of photoelectrons to a very high density of $10^{18}$ $cm^{-3}$ in that QD. Therefore, the density of mobile holes rather than that of the confined photoelectrons limits non-radiative intra-band relaxation rate of such photoelectrons in QDs. Furthermore, because of blocking barrier $\varepsilon_B$, the density of those mobile holes is proportional to $exp(-\varepsilon_B/kT)$ in the spacers. Hence, the blocking barrier reduces $exp(-\varepsilon_B/kT)$ times the intra-band relaxation rate in QDs and raises $exp(\varepsilon_B/kT)$ times the confined photoelectron lifetime.

## 2.2 Balance of currents

Figure 2 illustrates the energy band diagram of proposed IB solar cell and the electron flows. The red dashed vertical arrows denote $j_{VC}$, $j_{VI}$ and $j_{IC}$ electron flows generated by photon fluxes incoming from the $\varepsilon_{CV} < \hbar\omega$, $\varepsilon_{IV} < \hbar\omega < \varepsilon_{CI}$ and $\varepsilon_{CI} < \hbar\omega < \varepsilon_{CV}$ spectral ranges, respectively. The black dashed vertical arrows denote $j_{CV}$, $j_{IV}$ and $j_{CI}$ electron flows related to radiative recombination resulting in photon emission fluxes from the solar cell in the same spectral ranges, respectively. Approximating incoming and emitting photon fluxes by the Plank formula of blackbody emission and assuming complete absorption of incoming photons in each of those $[\varepsilon_i, \varepsilon_j]$ spectral ranges enable to write electron flows as $j_{ji}exp(\mu/kT)$,

$$j_{ji} = \frac{2eX}{h^3c^2GF}\int_{\varepsilon_i}^{\varepsilon_j}\frac{exp[(-\mu)/kT]\varepsilon^2 d\varepsilon}{exp[(\varepsilon-\mu)/kT]-1} \quad (1)$$

where $GF = 4.6\times10^4$ is the geometrical factor related to angle that Earth is seen from Sun, $X$ is the concentration of solar light, and $\mu = 0$ for photocurrents related to solar photon absorption. On the other hand, $X = GF = 1$ and $\mu$ is the splitting of quasi-Fermi levels for recombination currents related to photon emission from the solar cell.

The non-radiative recombination through QDs raises $j_{NIV}$ and $j_{NCI}$ electron flows denoted by black dashed vertical arrows in Figure 2. As noted above, the density of mobile holes in the stack determines non-radiative intra-band relaxation rate of confined photoelectrons in QDs. Therefore, the non-radiative recombination current from IB states into valence band reduces to

$$j_{NIV}[1 - exp(-\mu_I/kT)] \quad (2)$$

where $j_{NIV} = (ep_0 n_Q \Omega L/\tau_{ph})exp(-\varepsilon_B/kT)$ since the $\varepsilon_B$ barrier blocks the $p_0\Omega n_Q v_D$ flow of holes into the stack from the cap and buffer layers, $p_0$ is the concentration of holes in the cap and buffer layers; $n_Q$ is the density of QDs; $\Omega$ is the volume of QD; $L$ is the thickness of the stack; $\tau_{ph}$ is the intra-band relaxation lifetime mediated by optical phonons in QDs; $kT$ is the temperature of the cell ; and $\mu_I$ is the split between quasi-Fermi levels in QDs and valence band, respectively.

Assuming the same $F_C = eV + F_V$ quasi-Fermi level for conduction band electrons in n- and p-doped sides of p-n-junction, the current associated with electron-hole non-radiative recombination from conduction band into IB states can be written as

$$j_{NCI}[1 - exp\{(eV - \mu_I)/kT\}] \quad (3)$$

where $j_{NCI} = (en_i^2 L/p_0\tau_C)exp(\varepsilon_B/kT)$. Here $n_i$ is the intrinsic concentration of carriers for GaAs, $eV - \mu_I$ is the splitting in the quasi-Fermi levels of conduction band and QDs, and $\tau_C$ is the lifetime associated with non-radiative electron-hole inter-band recombination through QDs.

Balance $j_C = j_{VC} - j_{CV}$ of $j_{VC}$ and $j_{CV}$ electron flows yields conventional photocurrent as it is in the reference solar cell without QDs,

$$j_C = j_{SVC} - j_{CV}[exp(eV/kT) - 1] \quad (4)$$

where $V$ is the bias voltage.

Balance of $j_{VI}$, $j_{IV}$ and $j_{NIV}$ electron flows yields the net photocurrent into the confined electronic states from the valence band. By the same time, balance of $j_{IC}$, $j_{CI}$ and $j_{NCI}$ electron flows yields the net photocurrent from the confined electronic states into the conduction band. Obviously, both net currents are equal and they reduce to the additional photocurrent generated in IB solar cells,

$$j_Q = j_{SIC} - (j_{IC} + j_{NCI})\left(exp\frac{eV-\mu_I}{kT} - 1\right) = $$
$$= j_{SVI} - j_{VI}\left(exp\frac{\mu_I}{kT} - 1\right) - j_{NIV}\left(1 - exp\frac{-\mu_I}{kT}\right) \quad (5)$$

If $(\varepsilon_{IV} - \mu_I)/kT \gg 1$ and $(\varepsilon_{CI} + \mu_I - V)/kT \gg 1$, Equations (1) and (5) enable calculation of $\mu_I$ splitting as

$$exp\frac{\mu_I}{kT} = \frac{(j_{SVI} - j_{SIC} + j_{VI} - j_{IC} - j_{NIV} - j_{NCI})}{2j_{VI}}\times$$
$$\times\left[1 \pm \sqrt{1 + \frac{4j_{VI}\{(j_{IC}+j_{NCI})exp(eV/kT)+j_{NIV}\}}{(j_{SVI}-j_{SIC}+j_{VI}-j_{IC}+j_{NIV}-j_{NCI})^2}}\right] \quad (6)$$

Then the photocurrent $j = j_C + j_Q$ simply reduces to

$$j = j_{SVC} + j_{SIC} - j_{CV}[exp(eV/kT) - 1] - (j_{IC} + j_{NCI})\left(exp\frac{eV-\mu_I}{kT} - 1\right) \quad (7)$$

where $j_{NCI}\sim exp(\varepsilon_B/kT)$ and $j_{NIV}\sim exp(-\varepsilon_B/kT)$.

The net currents through QDs also bring charge into QDs for adjusting the $\varepsilon_B$ height of blocking barrier to that required by the balance of currents in the stack,

$$\varepsilon_B + \frac{4\sqrt{\varepsilon_B}(kT)^{3/2}}{e^2 l_D n_Q \Omega N_I L} + \frac{n_i^2 kT}{p_0 n_Q \Omega N_I}exp\frac{\varepsilon_B + eV}{kT} = \varepsilon_I - \mu_I - F_V \quad (8)$$

Sub-bandgap photons generate mobile photoelectrons that swiftly escape from conduction band of few *nm*-thick QDs into the conduction band of $Al_xGa_{1-x}As$ spacers and relax there in 1*ps*. The corresponding holes remain strongly confined in QDs. Since mobile holes are majority carriers in the spacers, they screen the escaped photoelectrons and participate in their ambipolar diffusion and drift driven by the pulling field of the graded stack. Such mobile photoelectron-hole pair can pass through the 500 *nm* grade stack by 50 *ps* while its inter-band recombination lifetime time is as long as 1-10 *ns* [7].

## 3 RESULTS AND DISCUSSION

### 3.1 Currents

Equations (5), (6) and (7) enable calculation of main photovoltaic parameters of proposed IB solar cell. These parameters are highly sensitive to the charge accumulated in QDs and spacers.

Equation (4) defines the photocurrent as function of bias applied to ideal p-n-junction solar cell. The shape of this I-V curve must be congruent with the dark current – voltage dependence that is the second term in the right hand side of Equation (4). Noteworthy such congruency is broken in GaSb/GaAs type-II QD IB solar cell because the charge accumulated in QDs modifies the shape of photocurrent curve (red line in Figure 3). The break is seen in Figure 3, where the blue curved line is the dark current of GaSb/GaAs type-II QD IB solar cell superimposed on the photocurrent.

Equation (7) defines the photocurrent of GaSb/GaAs QD IB solar cell. The first two terms in the right hand side of this equation comprise the short circuit current of the cell while the last two terms are generated by the bias. Exponential dependence on $\mu_I$ and $\varepsilon_B$ makes the bias-induced current highly sensitive to these parameters.

The splitting $\mu_I$ in quasi-Fermi levels describes modification of charge in QDs. Equation (5) shows that variation of $\mu_I$ impacts photocurrent while Equation (6) shows the dependence of $\mu_I$ on concentration of sunlight $X$ (via photocurrents $j_{SVC}$, $j_{SIC}$ and $j_{SVI}$) and material parameters $\tau_C$ and $\tau_{ph}$ of non-radiative inter-band recombination and intra-band relaxation lifetimes (via $j_{NCI}$ and $j_{NIV}$). Equation (6) also shows that the quasi-Fermi level $\mu_I$ is a function of both illumination and bias applied to IB solar cell.

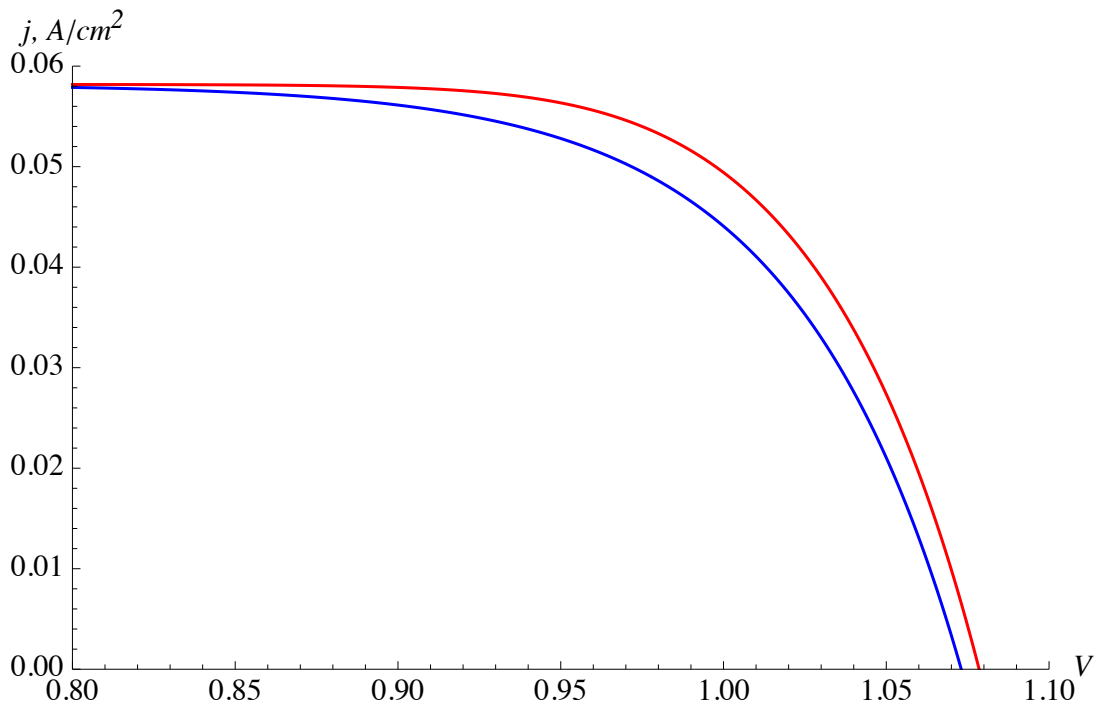

**Figure 3:** Current-voltage plot of GaSb/GaAs type-II QD IB solar cell for 1-sun concentration: blue curve is the dark current superimposed on photocurrent (red curve)

Figure 4 displays the difference of $\mu_I$ vs. bias dependences under 1sun illumination (red curve) and in the dark (blue curve). Though the difference of quasi-Fermi levels is large below *0.8V* bias, it does not impact congruency of curves in Figure 3 because the short circuit current component of Equation (7) dominates in both curves. Above *0.8V* the bias induced current raises so much that declines first the dark current curve then the photocurrent curve in Figure 3.

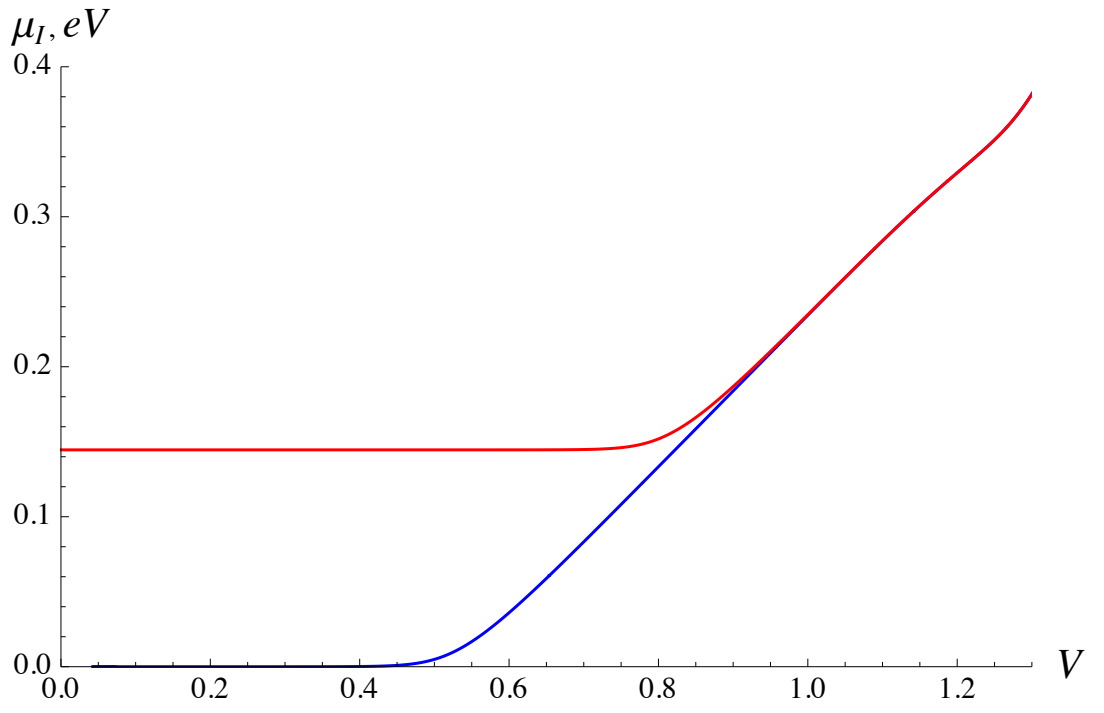

**Figure 4:** Quasi-Fermi level $\mu_I$ of photoelectrons confined in QDs of GaSb/GaAs type-II QDs IB solar cell as a function of bias under 1sun illumination (red curve) and in the dark (blue curve)

3.2 Concentration

Figure 5 displays the fill-factor of such solar cell as a function of sunlight concentration. The fill-factor *FF* of ideal GaSb/GaAs QD IB solar cell (dark red dots) is smaller than that of the ideal reference GaAs solar cell (black dots). The latter is about fixed at 91% and insensitive to sun-concentration while the fill-factor of proposed cell (dark red dots) swiftly grows to that value when concentration increases. For about 300-sun concentration, the fill-factor turns to the same saturation of 91%.

Figure 5 also shows that non-radiative electron transitions reduce the fill-factor of GaSb/GaAs QD IB solar cell. The faster the intra-band relaxation (red dots are for *1ps* and blue dots for *100ps)* of photoelectrons in QDs, the lower the saturation value is. Again the same 300-sun concentration is enough for turning of “non-radiative” fill-factor of such cells to saturation.

Our calculation of photovoltaic characteristics yields a linear dependence of photocurrent on concentration of sunlight for type-II GaSb/GaAs QD IB solar cell. For one sun concentration, the open circuit voltage $V_{OC}$ is as high as mainly determined by the GaAs p-n-junction. Otherwise the dependence of $V_{OC}$ on concentration is similar to that of the fill-factor as shown in Figure 6.

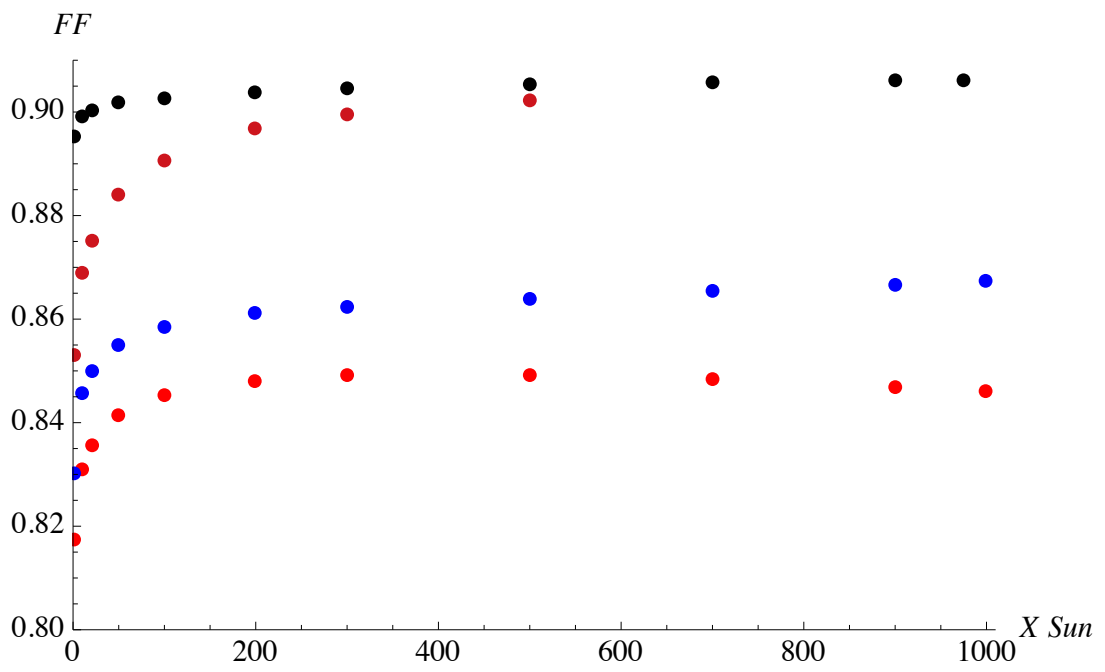

**Figure 5:** Fill-Factor FF of GaSb/GaAs QD IB solar cell as a function of sunlight concentration: radiative limit (dark red dots); *1ps* (red dots) and *100ps* (blue dots) non-radiative intra-band relaxation lifetime in QDs. Shockly-Queisser limit of reference GaAs solar cell (black dots)

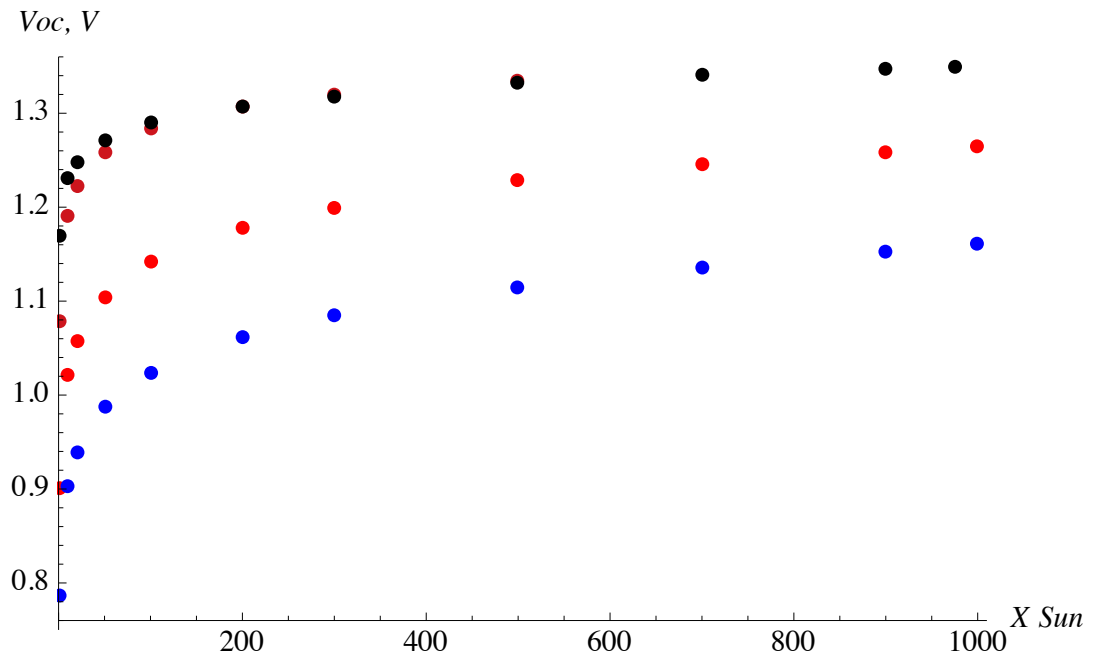

**Figure 6:** Open circuit voltage of GaSb/GaAs type-II QD IB solar cell as a function of sunlight concentration: non-radiative intra-band relaxation lifetime in QDs *1ps* (red dots) and *100ps* (blue dots); radiative limit (dark red dots). Shockly-Queisser limit of reference GaAs solar cell (black dots)

Our study shows that the clue to understanding of photovoltaic performance of GaSb/GaAs type-II QD IB solar cells is the accumulation of charge in QDs and spacers. The accumulated charge induces $\varepsilon_B$ screening barrier sandwiching the stack of QD layers. Since the absorption of sunlight modifies the charge, it also modifies the screening barrier. Figure 7 illustrates reduction of the barrier height by concentrated sunlight. Concentration of about 300-sun reduces the screening barrier $\varepsilon_B$ of ideal GaSb/GaAs QD IB solar cell (dark red dots) to the thermal energy of mobile carriers, *0.02eV*. In fact, such barrier disappears for mobile photoelectrons and holes. Hence, it cannot influence on photovoltaic performance of the cell if concentration is above 300-sun.

On the other hand non-radiative inter-band and intra-band transitions through QDs resist to modification of the charge already accumulated in QDs. Two processes included in Equation (5), photoelectron generation and their recombination through QDs, keep a balance of the charge. Quasi-Fermi level $\mu$ of photoelectrons confined in QDs regulates the accumulated charge. Non-radiative intra-band photoelectron relaxation in QDs reduces the splitting of quasi-Fermi level by *0.07eV* and *0.18eV* for *100ps* and *1ps* relaxation, respectively, as shown in Figure 8.

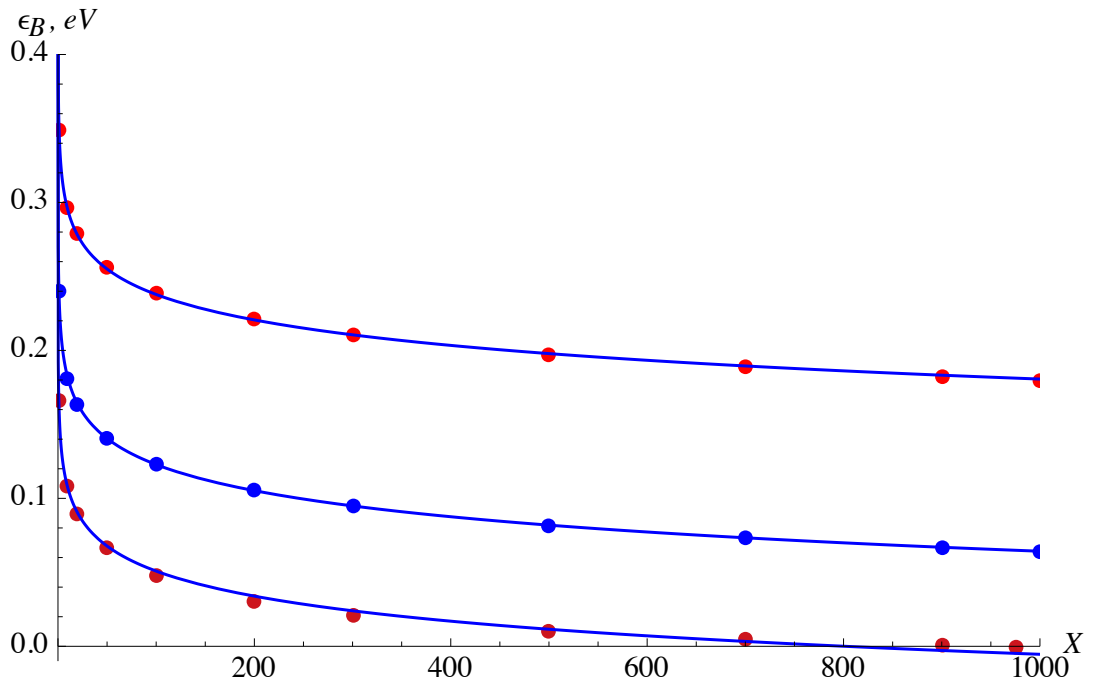

**Figure 7:** The screening barrier height $\varepsilon_B$ of GaSb/GaAs QD IB solar cell as a function of sunlight concentration: radiative limit (dark red dots); non-radiative intra-band relaxation lifetime in QDs *1ps* (red dots) and *100ps* (blue dots). Solid lines are $a + b \times ln\ (X)$ approximation of the height

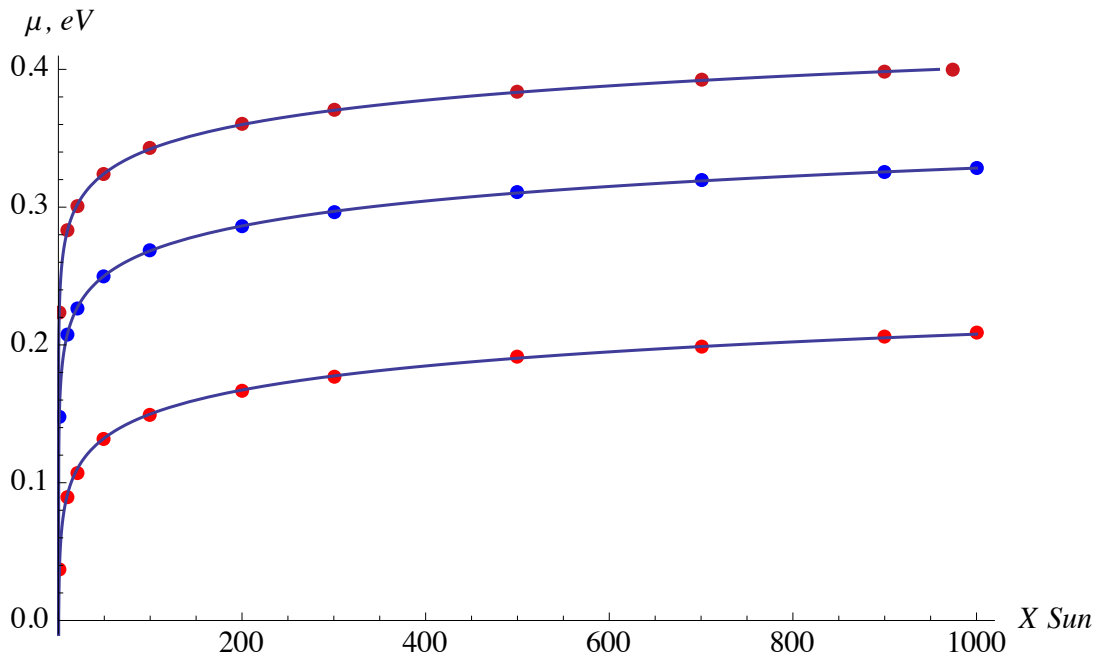

**Figure 8:** Quasi-Fermi level $\mu$ of photoelectrons confined in QDs in GaSb/GaAs QDs IB solar cell as a function of sunlight concentration: radiative limit (dark red dots); non-radiative intra-band relaxation lifetime in QDs *1ps* (red dots) and *100ps* (blue dots). Solid lines are $a + b \times ln\ (X)$ approximation of quasi-Fermi levels

3.3 Efficiency

Since photoelectrons modify the accumulated charge, concentration of sunlight is needed for achieving of higher performance.

Figure 9 displays the conversion efficiency $\eta$ of GaAs solar cells as a function of sunlight concentration. Solid lines are $a + b \times ln\,(X)$ approximation of the calculated efficiency. Black dots denote the Shockley-Queisser limit for the reference GaAs solar cell. The calculations show that two-photon absorption increases the efficiency of GaSb/GaAs QD IB cells. If only radiative recombination limits the cell performance (dark red dots), the efficiency achieves the highest value. Non-radiative electron transitions reduce the efficiency. For calculation, we use *1ps* (red dots) and *100ps* (blue dots) lifetime for non-radiative intra-band relaxation of confined photoelectrons in QDs since such intra-band transitions may be slower than *1ps* [8], and *10ns* lifetime for non-radiative inter-band recombination of conduction band mobile photoelectrons through QDs [7].

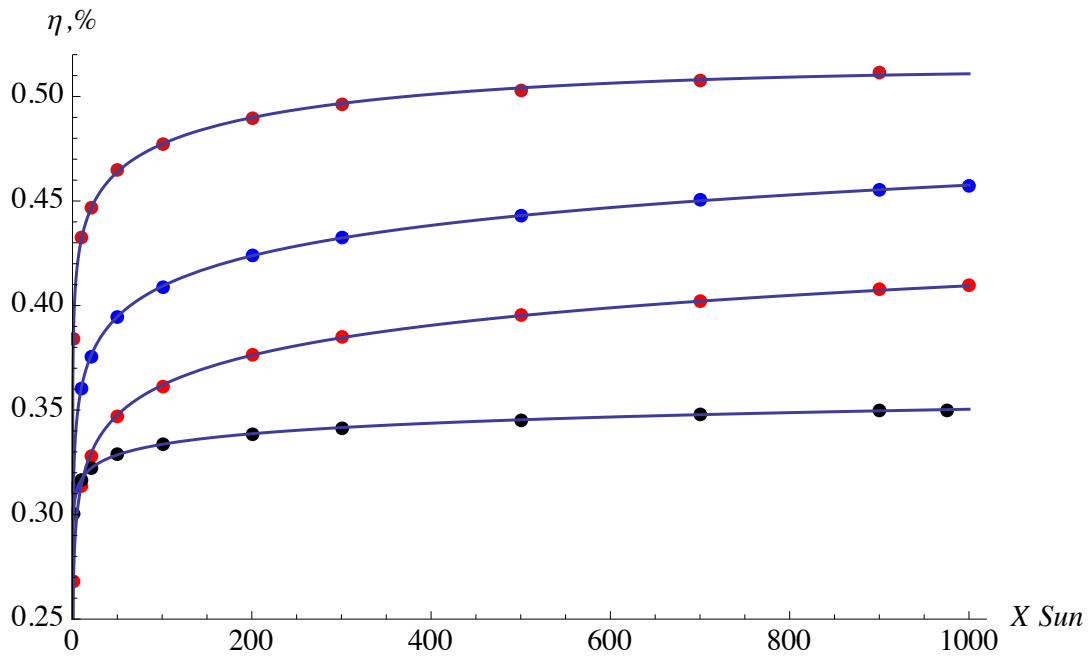

**Figure 9:** Conversion efficiency $\eta$ of GaSb/GaAs type-II QD IB solar cell as a function of sunlight concentration: radiative limit (dark red dots); non-radiative intra-band relaxation lifetime in QDs *1ps* (red dots) and *100ps* (blue dots). Shockly-Queisser limit of reference GaAs solar cell (black dots). Solid lines are $a + b \times ln\,(X)$ approximation of efficiency

Our calculation shows that concentration of sunlight from 1-sun to 500-sun raise the efficiency of proposed GaSb/GaAs QD IB solar cell from 30% to 50%. Further concentration of light has little effect. Though non-radiative relaxation of photoelectrons confined in QDs degrades the efficiency, it is still above the Shockley-Queisser limit by 5% to 10%.

It should be noted that the Luque-Marti limit of IB GaAs solar cells, 60% for 46000-sun concentration [9], is higher than the best efficiency shown in Figure 9 for GaSb/GaAs QD IB solar cell, 50% for 500-sun, because of the difference in concentration. In fact, the limiting performance is the same for both devices. For instance, the Luque-Marti limit is 51.6% for 1000-sun concentration [10].

## 4 CONCLUSIONS

In conclusion our study shows that QDs may help generation of the additional photocurrent by resonant two-photon absorption of sunlight in type-II GaSb/GaAs QD IB solar cell. In the proposed cell design QDs are located outside the depletion region. Special caution should be taken here because, like artificial atoms, QDs may easily convert their confined ground electronic state into fast recombination level. Such conversion of QD states degrades both additional and conventional photocurrent of IB solar cells because confined states promptly achieve equilibrium with conduction or valence bands.

Our calculation shows that concentration of sunlight enforces two-photon absorption in proposed GaSb/GaAs QD IB solar cell. It also reduces recombination through the “outside” QDs, which is important for achieving high conversion efficiency. Concentration from 1-sun to 500-sun raises the efficiency of proposed GaSb/GaAs QD IB solar cell from 30% to 50%. We showed that the clue to understanding of performance of GaSb/GaAs QD IB solar cells is the accumulation of charge in QDs and spacers.

To take into account non-ideality of real structures, we also include into consideration non-radiative intra-band relaxation of confined photocarriers in QDs and inter-band recombination of mobile carriers through QDs.

Non-radiative electron transitions through QDs enforce degradation of photocurrents, however, concentration of sunlight can reduce degradation in case of “outside” QDs. Our calculation shows that due to accumulation of charge in “outside” QDs, the efficiency of IB solar cell raises by 12%, from 27% for 1-sun concentration to 39% for 500-sun, as shown in Figure 9. The same concentration improves efficiency of ideal GaAs solar cell by 4% only, from 30% to 34%.

An important advantage of proposed GaSb/GaAs QD IB solar cell is that QDs are located outside of depletion region, at far enough distance so that they cannot cause current leakage through the depletion region. Thermal assistance or solar photons must be involved for confined electrons may escape from such QDs.

Recombination through QDs is a major factor that limits efficiency of solar cells based on QDs buried within the depletion region. Our proposal will help to solve this problem.

## ACKNOWLEDGEMENTS

A.Kechiantz and A. Afanasev acknowledge support from The George Washington University. A. Kechiantz and J.-L. Lazzari acknowledge support from the bilateral Armenian/French project, SCS/CNRS contract #IE-013/ #23545.